# A Hybrid Deep Learning CNN Model for Enhanced COVID-19 Detection from Computed Tomography (CT) Scan Images

**Suresh Babu Nettur[1*], Shanthi Karpurapu[1*], Unnati Nettur[2], Likhit Sagar Gajja[3], Sravanthy Myneni[1], Akhil Dusi[4] AND Lalithya Posham[5]**

[1] Independent Researcher, Virginia Beach, VA 23456, USA
[3] Department of Computer Science, Virginia Tech, Blacksburg, VA 24061 USA
[3] Department of Computer Science, BML Munjal University, Haryana 122413 INDIA
[4] Department of Information Systems, University of Indiana Tech, Indiana USA
[5] Nanjing Medical University, Nanjing City, Jiyangsu, CHINA

Corresponding authors: Shanthi Karpurapu (shanthi.karpurapu@gmail.com), Suresh Babu Nettur (nettursuresh@gmail.com)

[*] Shanthi Karpurapu and Suresh Babu Nettur are co-first authors

**ABSTRACT** Early detection of COVID-19 is crucial for effective treatment and controlling its spread. This study proposes a novel hybrid deep learning model for detecting COVID-19 from CT scan images, designed to assist overburdened medical professionals. Our proposed model leverages the strengths of VGG16, DenseNet121, and MobileNetV2 to extract features, followed by Principal Component Analysis (PCA) for dimensionality reduction, after which the features are stacked and classified using a Support Vector Classifier (SVC). We conducted comparative analysis between the proposed hybrid model and individual pre-trained CNN models, using a dataset of 2,108 training images and 373 test images comprising both COVID-positive and non-COVID images. Our proposed hybrid model achieved an accuracy of 98.93%, outperforming the individual models in terms of precision, recall, F1 scores, and ROC curve performance.

## I. INTRODUCTION

In March 2020, the World Health Organization (WHO) declared COVID-19, caused by the SARS-CoV-2 virus, a global pandemic. COVID-19 is highly contagious and can develop into severe acute respiratory distress syndrome (ARDS), which can be life-threatening. Early detection and diagnosis are critical for controlling the spread of the virus. The reverse-transcription polymerase chain reaction (RT-PCR) test is the most commonly used method for COVID-19 screening. However, this technique is time-consuming, and studies have noted that its sensitivity is low in the early stages of infection [1]. Chest imaging techniques, including X-rays and computed tomography (CT) scans, have also been used to detect lung abnormalities associated with COVID-19. However, the accuracy of COVID-19 diagnosis using chest imaging relies significantly on the expertise of radiologists [2]. Recently, several studies have investigated deep learning techniques as a tool to aid in and automate the diagnostic process [3] [4] [5] [6] [7] [8][9] [10] [11] [12] [13] [14] [15] [16] [17] [18] [19] [20] [21].

A CT scan generates detailed images of organs, bones, soft tissues, and blood vessels, providing physicians with critical insights into internal structures, including their shape, size, density, and texture. Unlike conventional X-rays, which overlay different structures in a single image, CT scans produce a series of cross-sectional "slices" of a specific body region, offering a more comprehensive view. This level of detail aids in accurately identifying potential medical issues, along with their exact location and severity. Consequently, numerous deep learning-based methods have recently been proposed for COVID-19 screening using CT scan images [22] [23] [24] [25] [26] [27].

The overarching goal of our research is to advance the application of deep learning and LLMs across software engineering [28] [29] and bio medical domains [30] [31]. In light of recent research advancements, our primary objective of this research is to develop a high-performing deep learning-based classification system to enhance the detection and diagnosis of COVID-19 using CT scan images. Our study seeks to address critical challenges in existing approaches, such as ineffective feature extraction, computational inefficiencies, and the risk of overfitting. By integrating powerful pre-trained models with advanced dimensionality reduction techniques, our research aims to achieve superior performing model in distinguishing COVID-19 cases from non-COVID-19 cases.

## II. RELATED WORKS

Since the COVID-19 outbreak, researchers have increasingly focused on creating deep learning methods to screen the disease using medical imaging techniques such as CT scans and chest X-rays. We specifically explored previous research focused on deep learning methods using CT scans for COVID-19 detection, as our approach also utilizes CT scan images. Given our focus on CT-based COVID-19 detection, we analyzed prior research utilizing deep-learning approaches related to CT imaging to enhance our methodology.

### A. CNN AND TRANSFER LEARNING APPROACHES
Several studies have focused on leveraging CNNs, particularly transfer learning, for COVID-19 detection using CT scans. Xu et al. (2019) [32] developed a novel deep learning method using a location-attention mechanism and ResNet architecture to automatically screen COVID-19 CT images in this multi-center case study. The model achieved an 86.7% accuracy in classifying COVID-19, IAVP, and healthy cases, showing promise as a supplementary diagnostic tool for frontline clinicians [32]. He et al. (2020) [22] developed the COVID-19 CT dataset with 349 CT scans and proposed the Self-Trans approach, combining self-supervised learning with transfer learning. Their method achieved an F1 score of 0.85 and AUC of 0.94, demonstrating high accuracy in diagnosing COVID-19 with limited data [22]. Wang et al. (2021) [26] proposed an artificial intelligence-based method using a modified Inception transfer-learning model to diagnose COVID-19 from CT images, achieving 89.5% accuracy in internal validation and 79.3% in external validation [26]. Amyar et al. (2020) [25] proposed a multi-task deep learning model for simultaneously classifying COVID-19 and segmenting lesions in chest CT images.

The model, which jointly performs segmentation, classification, and reconstruction tasks, achieved a dice coefficient higher than 0.88 for segmentation and an AUC of 0.97 for classification [25]. Wang et al. (2020) [33] introduced a joint learning framework to improve COVID-19 CT diagnosis by learning from heterogeneous datasets. They redesigned COVID-Net, incorporating feature normalization and a contrastive training objective, achieving 12.16% and 14.23% higher AUC than the original model on two large-scale datasets, outperforming other multi-site learning methods [33]. Han et al. (2020) [34] introduced AD3D-MIL, a model for weakly-supervised COVID-19 screening from chest CT, achieving 97.9% accuracy, 99.0% AUC, and 95.7% Cohen kappa. The model uses attention-based pooling for high accuracy and interpretability, showing strong potential as an efficient tool for large-scale COVID-19 screening [34]. Polsinelli et al. (2020) [24] developed a light CNN model based on SqueezeNet for distinguishing COVID-19 CT images. CNN-2 achieved 85.03% accuracy, 87.55% sensitivity, and 86.20% F1-score. It classified images in 1.25 seconds on a high-end workstation and 7.81 seconds on a medium laptop without GPU acceleration.

Performance can be improved with efficient pre-processing [24]. Haryanto et al. (2024) [35] proposed SCOV-CNN, a convolutional neural network for COVID-19 classification based on CT images. Inspired by LeNet, it uses a deeper architecture with seven and five kernel sizes and three fully connected layers with dropout. Evaluated on CT images from 120 patients, SCOV-CNN achieved 96% accuracy, 98% precision, and 95% F1 score [35].

### B. EXPLAINABLE AI APPROACHES
Some recent works have focused on explainability and improving clinical acceptance through transparent AI. Soares et al. (2020) [27] introduced a dataset with 2482 CT scans (1252 COVID-19 positive, 1230 negative) collected from São Paulo, Brazil. They applied the xDNN classifier, achieving an F1 score of 97.31%. The xDNN model provides explainable results with IF...THEN rules for early diagnosis [27]. Rajpoot et al. (2024) [36] proposed an ensemble approach combining CNN models with explainable AI techniques (LIME, SHAP, Grad-CAM, Grad-CAM++), achieving high. Their work emphasizes model transparency and interpretability, bridging the gap between precision and clinical applicability [36]. This approach ensures that deep learning models for COVID-19 detection are not only accurate but also interpretable by clinicians, increasing their trust and adoption in clinical settings.

### C. HYBRID AND ENSEMBLE APPROACHES

Other research has focused on hybrid and ensemble models that combine multiple techniques for improved performance. Mobiny et al. (2020) [23] introduced Detail-Oriented Capsule Networks (DECAPS) for automatic COVID-19 diagnosis from CT scans. DECAPS integrates Capsule Networks with enhancements like Inverted Dynamic Routing, Peekaboo training, and data augmentation using generative adversarial networks. The model achieves 84.3% precision, 91.5% recall, and 96.1% AUC, outperforming state-of-the-art methods and experienced radiologists, suggesting its potential to assist in CT scan-based COVID-19 diagnosis [23]. Islam et al. (2022) [37] proposed an ensemble model for COVID-19 CT image classification, addressing the limitations of RT-PCR by using CT scans for detection. They applied contrast-limited histogram equalization (CLAHE) for image enhancement and developed a Convolutional Neural Network (CNN).

The extracted features were used with various machine learning algorithms-Gaussian Naive Bayes (GNB), Support Vector Machine (SVM), Logistic Regression (LR), Decision Tree (DT), and Random Forest (RF). The ensemble model outperformed state-of-the-art models [37]. Kundu et al. (2022) [38] developed an ensemble-based framework called ET-NET for automated COVID-19 detection using chest CT-scan images. Their approach employs a bootstrap aggregating (bagging) technique, integrating three transfer learning models-Inception v3, ResNet34, and DenseNet201-to enhance classification performance, achieving an impressive accuracy of 97.73% [38]. Aversano et al. (2021) [39] introduced an ensemble-based approach for COVID-19 detection using CT scan images. By combining pre-trained networks (VGG, Xception, ResNet) optimized via a genetic algorithm, the method classifies clustered lung lobe images using a majority voting strategy. The ensemble outperformed single classifiers, achieving F1-scores of 0.94–0.95 on an integrated dataset, demonstrating improved generalization and stability across diagnostic contexts [39]. Shaik et al. (2022) [40] proposed an ensemble-based approach for detecting COVID-19 infection from chest CT scan images, aggregating predictions from multiple fine-tuned pre-trained models such as VGG16, InceptionV3, ResNet50, Xception, and MobileNet. Their method leverages a composite ensemble classifier that combines candidate model predictions, achieving superior results [40]. Maftouni et al. (2021) [41] developed an ensemble model for COVID-19 diagnosis using chest CT scans, combining Residual Attention-92 and DenseNet-121 to leverage complementary features. A meta-learner integrates the outputs of these networks, achieving superior performance with an accuracy of 95.07% and ROC AUC of 96.72% [41]. De Jesus Silva et al. (2023) [42] proposed four ensemble CNN models using transfer learning for COVID-19 detection from CT scans and compared them with state-of-the-art CNN architectures. After testing 11 models, they selected DenseNet169, VGG16, and Xception. The ensemble of these three models, called EnsembleDVX, achieved the best results with an accuracy of 97.7%, precision of 97.7%, recall of 97.8%, and an F1 score of 97.7% [38].

Our approach uniquely combines transfer learning using DenseNet121, VGG16, and MobileNetV2, followed by feature extraction, dimensionality reduction with PCA, and classification using SVC. Our method significantly improves accuracy and AUC scores for distinguishing between COVID-19 and non-COVID-19 cases, which sets our approach apart from the studies reviewed. Additionally, dimensionality reduction techniques, such as PCA, are often not adequately integrated into these systems, leading to inefficient feature representation and model performance degradation. Furthermore, our approach uniquely tackles the challenge of dimensionality reduction through PCA, effectively minimizing computational costs and reducing the risk of overfitting, issues that are prevalent in many existing models, particularly in the context of COVID-19 classification.

### III. METHODOLOGY
The proposed hybrid deep learning model is presented in Figure 1. We first normalized the images to meet pre-trained CNN model input requirements and applied image augmentation to enhance dataset diversity and reduce overfitting. Next, to leverage the strengths of deep learning, we adopted a transfer learning approach using three pre-trained CNNs: MobileNetV2, DenseNet121, and VGG16. These networks were employed to extract deep features from the processed CT scan images. This step is vital for effectively capturing and representing both high-level and low-level features from CT scan images, ensuring that critical image details are represented for accurate analysis and classification. Following this, to handle the high dimensionality of the extracted features, we applied PCA to transform the features into a lower-dimensional space, retaining the most significant variations in the data. This process reduced redundancy and noise, improved computational efficiency, and ensured that the most discriminative information was preserved for downstream classification. The reduced features from all three pre-trained networks were then stacked (concatenated) together to form a final unified feature set, combining the diverse and complementary information captured by each model. Finally, the stacked features were fed into SVC to train the model and perform the final classification, enabling the effective detection of COVID-19. In this section, we will outline the key components of the methodology.

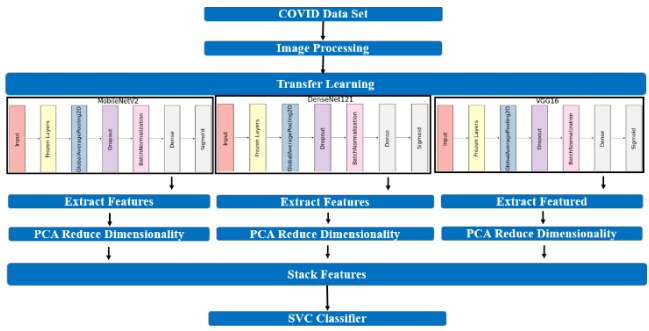

**FIGURE 1.** Hybrid Deep Learning Model Approach

### A. DATASET

We used the SARS-CoV-2 CT scan dataset available on Kaggle [43] (PlamenEduardo, 2020), originally collected by Angelov and Almeida Soares [27] (2020) from hospitals in São Paulo, Brazil. The SARS-CoV-2 CT-scan dataset consists of 2481 CT scans from 120 patients, with 1252 CT scans of 60 patients infected by SARS-CoV-2 from males (32) and females (28), and 1229 CT scan images of 60 non-infected patients by SARS-CoV-2 from males (30) and females (30), but presenting other pulmonary diseases. Data was collected from hospitals in São Paulo, Brazil. The dataset includes CT images with varying sizes, ranging from 182×129 pixels for the smallest images to 484×416 pixels for the largest. Some examples of these images are shown in Figure 2. The dataset was split into two sections: 85% of the images were used for training, and 15% were reserved for testing to facilitate model training and evaluation. We chose this dataset as it is from real-time patients collected from multiple hospitals in São Paulo, Brazil [27]. The dataset's diversity of patient cases and image sizes, along with its prior testing with various methods [27], makes it a well-established resource for evaluating COVID-19 detection models. As a next step, image pre-processing and augmentation techniques were applied to enhance the quality and variability of the dataset, ensuring its suitability for efficient model training and evaluation.

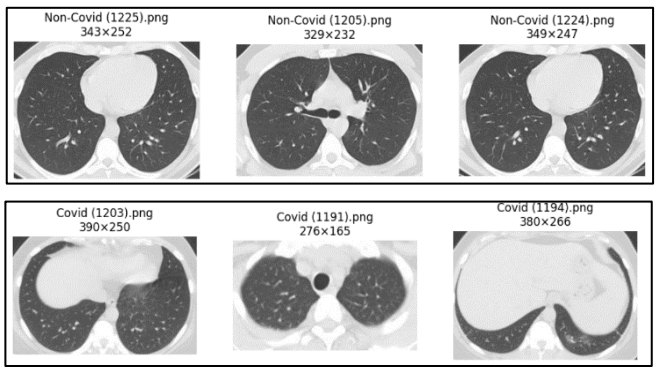

**FIGURE 2.** Sample COVID, non-COVID Images from Dataset

### B. IMAGE PRE-PROCESSING AND AUGMENTATION

In computer vision tasks, image pre-processing is a crucial step in preparing the data for model training. The pre-processing technique helps improve model performance by addressing various factors such as noise reduction, image normalization, and resizing. These steps are particularly important for CNNs, which rely on consistent input dimensions and well-scaled pixel values. In our work, pixel intensity normalization is applied to scale the pixel values to the range of [0, 1]. This normalization step is essential for stabilizing the training process and ensuring efficient model convergence, enabling the model to learn patterns more effectively without being influenced by variations in image brightness or contrast. Additionally, we resized the images to ensure compatibility with the network's input dimensions. This resizing process is crucial for maintaining consistency across the dataset, especially when using pre-trained models that require fixed input sizes. The choice of 224x224 pixels aligns with common practices in the field, where such dimensions are frequently used in models like MobileNetV2, DenseNet121, and VGG16, ensuring efficient training and inference.

We also performed image augmentation to enhance dataset diversity by simulating real-world variances, which helps in the development of a more resilient model and reduces overfitting. Our approach involved configuring a range of transformations, including rotation, width and height shifts, shear, zoom, and brightness adjustments. For instance, slight rotations and shifts were applied to allow the model to recognize features from different perspectives, while brightness and zoom adjustments helped the model generalize across varied lighting conditions and scales. These augmentations ensure that the model can learn more robust features, improving its ability to handle a variety of real-world scenarios and data variations. Table 1 provides a detailed overview of the augmentation techniques, specifying the parameters used for each image transformation. As the next step, the prepared and augmented dataset is used to apply transfer learning for fine-tuning pre-trained CNN models to detect COVID.

TABLE I
IMAGE AUGMENTATION PARAMETER VALUES

| Augmentation Parameter | Value | Description |
|---|---|---|
| Rotation Range | ±10 degrees | Randomly rotates images within ±10 degrees |
| Width Shift Range | 5% | Shifts images horizontally by up to 5% of the image width |
| Height Shift Range | 5% | Shifts images vertically by up to 5% of the image height |
| Shear Range | 0.1 | Applies a random shear transformation with intensity of 0.1 |
| Zoom Range | 10% | Randomly zooms in or out by up to 10% |
| Brightness Range | [0.9, 1.1] | Randomly adjusts brightness within the specified range |
| Fill Mode | Reflect | Fills points outside the boundaries by reflecting edges |

### C. TRANSFER LEARNING

We adopted a transfer learning approach to fine-tune pre-trained CNN models, such as MobileNetV2, VGG16, and DenseNet121, for COVID-19 detection. These models were initially trained on ImageNet, a large-scale dataset containing 1.28 million natural images divided into 1,000 categories. By leveraging feature representations learned from large datasets like ImageNet, this strategy significantly reduced the need for extensive labeled data, which is often scarce during pandemics. It also enabled rapid adaptation to the specific task while enhancing diagnostic accuracy. Additionally, the approach minimized computational demands, making it particularly effective in resource-constrained settings. Transfer learning thus provides a scalable and cost-effective solution to improve COVID-19 detection, addressing challenges of data scarcity and limited resources. The CNN models chosen for this study, MobileNetV2, DenseNet121, and VGG16, were selected based on state-of-the-art review and are discussed in detail in this section.

### D. MOBILENETV2

MobileNetV2, introduced by Sandler et al. (2018) [44], is a CNN architecture optimized for mobile and embedded vision applications. It uses an inverted residual structure with shortcut connections between compact bottleneck layers, which helps reduce the number of parameters and improve computational efficiency. The design begins with a 32-filter convolutional layer, followed by 19 bottleneck layers, which allow for deeper networks while minimizing the size of intermediate layers. MobileNetV2 is well-suited for tasks requiring efficient performance, such as object detection, image segmentation, and real-time inference on mobile and edge devices [45] [46] [47] [48] [49] [50].

### E. DENSENET121

DenseNet121, introduced by Huang et al. (2017) [51], consists of a dense connectivity design where each layer is directly connected to all its preceding layers. This architecture features a dense connection pattern that mitigates the vanishing gradient problem during the training of deeper architectures. Each layer has direct access to both the gradients from the loss function and the original input signal, hence enhancing the flow of gradients and information throughout the network. It also ensures the network is more compact and efficient. The architecture of DenseNet121 is built on dense blocks that include several convolutional layers, batch normalization units, and dense connections. To further minimize the dimensionality of feature maps, DenseNet121 incorporates transition layers. These layers utilize 1×1 convolutions and average pooling to compress the size of feature maps, thereby reducing computational complexities. DenseNet121 achieved remarkable performance in the ImageNet classification challenge and is widely utilized across various computer vision applications, including object detection and image segmentation [52] [53] [54] [55] [56] [57]. The design principle of dense connectivity offers essential insights for developing efficient and accurate deep neural networks. It captures complex, high-level features effectively, making it a key component for tasks requiring detailed feature extraction, such as disease detection.

### F. VGG16

VGG16, introduced by Simonyan and Zisserman in 2014 [58], is a well-known convolutional neural network (CNN) architecture recognized for its simplicity and effectiveness. The model increases depth by stacking small 3x3 convolution filters, allowing it to capture intricate patterns in images without requiring complex operations. VGG16 consists of 16 weight layers, including 13 convolutional layers and three fully connected layers. The architecture starts with two convolutional layers (64 filters), followed by max pooling. This pattern is repeated, with the number of filters increasing at each layer (e.g., 128 filters in Conv_2, 256 filters in Conv_3, and 512 filters in Conv_4 and Conv_5), each followed by max pooling. The network concludes with three fully connected layers and a Softmax activation function. VGG16 excels at extracting low-level features such as edges and textures, making it highly effective for capturing these fundamental patterns in images. Trained on the ImageNet dataset, VGG16 has become a foundational model in deep learning and computer vision. It provides a strong baseline for extracting essential image features, making it a valuable complement to the strengths of other architectures in a hybrid approach. Numerous follow-up research studies have demonstrated the model's utility and flexibility, leading it to be a foundational model in deep learning and computer vision research [59] [60] [61] [62] [63] [64] [65] [66].

### G. FINE-TUNING PRE-TRAINED MODELS

After selecting the three pre-trained CNN architectures, we implemented a fine-tuning strategy tailored to COVID-19 detection for the models. By leveraging pre-trained CNN models, we retained several of their initial layers, which were frozen to preserve the general features learned from ImageNet.

Freezing these layers means that their weights were not updated during training on the new dataset. This approach prevents the model from overwriting the general-purpose feature representations, such as edges, textures, and basic shapes, that these layers have already learned from the large and diverse ImageNet dataset. The final classification layer is replaced with new layers customized for the COVID-19 detection task. Only the newly added layers were left unfrozen, meaning their weights were updated during training. This allowed for targeted fine-tuning that enhanced task-specific performance by adapting to COVID-19 detection while maintaining computational efficiency. The transfer learning process we followed is illustrated in Figure 3.

As shown in Figure 3, the custom layers added include a global average pooling layer to summarize the features. This layer helps reduce the spatial dimensions of the feature maps, offering computational efficiency and preventing overfitting by generating compact representations [67]. A dropout layer is included to combat overfitting by randomly omitting a fraction of the units during training. Dropout acts as a regularizer by forcing the network to rely on a subset of neurons, which is shown to enhance generalization by simulating a bagged ensemble of neural networks [68]. Furthermore, a batch normalization layer is used to improve the stability and speed of training by normalizing the output of each layer, ensuring zero mean and unit variance [69]. This regularization technique helps accelerate convergence while stabilizing learning. The ReLU (Rectified Linear Unit) activation function is applied to introduce non-linearity. ReLU, defined as $f(x)=\max(0,x)$, has become one of the most popular activation functions due to its simplicity and efficiency, promoting sparsity and mitigating the vanishing gradient problem. The final dense layer uses a sigmoid activation function to produce the probabilities for the binary classification task. The sigmoid function, as defined in Equation 1, transforms inputs into a value between 0 and 1, effectively representing the probability of the positive class in a binary classification task.

$$\sigma(x) = \frac{1}{(1+e^{-x})} \qquad (1)$$

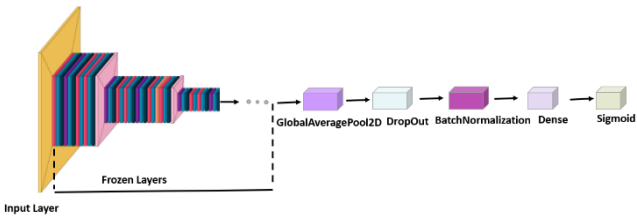

**FIGURE 3.** Transfer Learning Architecture

We present a detailed comparison of the parameter configurations in three popular architectures, DenseNet121, VGG16, and MobileNetV2, with and without a transfer learning approach in Table 2. We outline two scenarios for each model: training all layers versus training only the newly added layers. In the first scenario, where all layers are trained, the total, trainable, and non-trainable parameters are documented for each architecture. In the second scenario, where only the newly added layers are trained, we highlight the significant reduction in the number of trainable parameters achieved by the transfer learning strategy. This reduction factor demonstrates the computational efficiency of the models, enabling faster training with fewer resources while maintaining high performance. The input shape used for all models is (224, 224, 3). Finally, the fine-tuned models are used for feature extraction in the proposed model, and their features are combined and fed into a classification model. The reduction in trainable parameters for individual models through transfer learning is shown in Figure 4.

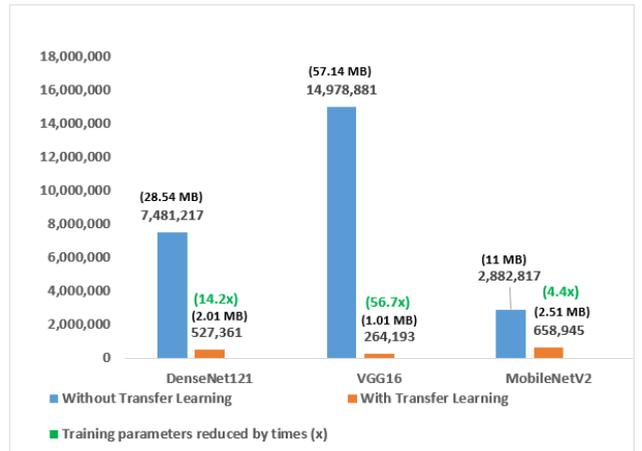

**FIGURE 4.** Transfer Learning Architecture

### H. ENSEMBLE LEARNING

Ensemble learning, which merges the predictions of several classifiers, has emerged as a robust strategy for image classification, often yielding higher performance than individual classifiers. In the context of image classification, ensemble techniques typically involve combining the outputs of various base classifiers, such as support vector machines, decision trees, and neural networks. These models leverage the unique characteristics of each classifier, enhancing accuracy and robustness by capturing different features of the data while compensating for their respective strengths and weaknesses. Common methods used to create ensemble classifiers include bagging, boosting, and stacking. Bagging, or Bootstrap Aggregating, generates diverse models by training each classifier on a bootstrapped subset of the training dataset. On the other hand, boosting improves weak learners iteratively by focusing more on misclassified examples, thereby enhancing their performance over successive iterations. Stacking employs a meta-classifier that learns to combine the predictions of base classifiers, using these outputs as inputs to generate a higher-level representation of the data. By integrating multiple classifiers, ensemble learning offers a promising approach to advancing the state-of-the-art in image classification. This method effectively combines the outputs of different models, improving overall performance and robustness in complex classification tasks.

*I. PROPOSED HYBRID MODEL*
In our research, we employed a stacking approach to build the hybrid model. This method is particularly effective, as each CNN can extract distinct sets of image features, reducing the potential loss of important information and improving the overall representation. In CNNs, the initial layers capture simple shapes, while the deeper layers identify complex, high-level features, making the fusion of features from different CNN models highly beneficial for enhanced performance. Before employing the stacking technique, we performed feature extraction from the three models: DenseNet121, VGG16, and MobileNetV2. Specifically, the final dense layer (the layer just before the output layer) of each model was utilized as a feature extractor. This approach captured intricate patterns and characteristics from the images. By leveraging this dense layer, we effectively transferred learned features to enhance the COVID-19 classification task.

The stacking of features from multiple models leads to an increase in the dimensionality of the feature space, which can introduce redundancy and computational challenges. To address these issues, after feature extraction, we applied PCA to reduce the dimensionality of the extracted features and optimize them for classification tasks. PCA is a statistical technique that transforms the original feature set into a smaller set of uncorrelated components, known as principal components, which capture the most variance in the data. Prior to applying PCA, the features were standardized to ensure uniform scaling. This step was crucial, as it ensured that each feature contributed equally to the PCA analysis, preventing features with larger numerical ranges from dominating the results. Standardization was performed by subtracting the mean and scaling the features to have unit variance. After this step, we applied PCA to reduce the dimensionality of the data while preserving as much variance as possible, ensuring a more meaningful and efficient representation of the features. Its linear nature also makes it computationally efficient. To decide how many principal components to retain, we performed an explained variance analysis with the goal of capturing at least 95% of the total variance. This approach strikes a balance between reducing dimensionality effectively and preserving crucial information, ultimately improving the performance of the COVID-19 classification model.

After this, we employed a feature-level stacking approach, where features extracted from all three models were concatenated to form a comprehensive feature set. These combined features were then used to train the Support Vector Classification (SVC) model, which classifies the features by finding the optimal hyperplane that best separates the data. The kernel function in SVC maps the feature space into a higher-dimensional space, allowing the model to learn from the diverse representations captured by each architecture. The approach aimed to leverage the strengths of each model's feature extraction capabilities, thereby enhancing overall classification performance for distinguishing between COVID-19 and non-COVID-19 cases.

*J. EVALUATION*
We evaluated the results of a direct transfer learning approach, which involves using pre-trained models directly without additional processing, alongside our proposed hybrid deep learning model. To rigorously assess all the models' effectiveness in binary classification for COVID-19 diagnosis, we employed several key evaluation metrics: accuracy, precision, recall, F1 score, the area under the ROC curve (AUC-ROC), and the confusion matrix. These metrics comprehensively analyzed each model's ability to distinguish between COVID-19-positive and normal cases, offering insights into class-specific performance and overall classification strength. This thorough evaluation enabled us to clearly compare the direct transfer learning models and our proposed hybrid deep learning model, underscoring the advantages of our technique in leveraging diverse feature representations for robust COVID-19 detection.

*K. ACCURACY*
Accuracy is the simplest evaluation metric, calculated as the ratio of correct predictions to the total predictions, providing an overall measure of each model's performance (Equation 2). It is represented as TP representing True Positives, TN representing True Negatives, FP representing False Positives, and FN representing False Negatives.

$$Accuracy = \frac{TP+TN}{TP+TN+FP+FN} \qquad (2)$$

*L. CONFUSION MATRIX*
The confusion matrix offers a deeper view of the model's performance across each class by providing counts of True Positives, True Negatives, False Positives, and False Negatives. This metric helps evaluate class-level performance and diagnose potential class imbalances or misclassifications.

*M. PRECISION, RECALL, AND F1 SCORE*
Precision, Recall, and F1 Score offer class-specific insights, particularly for evaluating model performance on imbalanced datasets. These metrics are derived from the confusion matrix values. Precision (Equation 3) is defined as the ratio of correctly predicted positive cases (True Positives) to all predicted positives, indicating the accuracy of positive predictions. Recall (Equation 4) is defined as the ratio of correctly predicted positives to all actual positives, capturing the model's ability to detect COVID-19 cases. The F1 (Equation 5) score is defined as the harmonic mean of Precision and Recall, providing a balanced measure of performance for each class.

$$Precision = \frac{TP}{TP+FP} \quad (3)$$

$$Recall = \frac{TP}{TP+FN} \quad (4)$$

$$F1 = 2 \times \frac{Precision \times Recall}{Precision + Recall} \quad (5)$$

### N. WEIGHTED AVERAGE

In classification performance metrics, weighted averages are used to summarize Precision, Recall, and F1 scores across multiple classes, particularly in imbalanced datasets. These averages provide insights into model performance by considering both class imbalance and individual class performance, enhancing the interpretation of results beyond per-class metrics. The weighted average (Equation 6) is a weighted mean of the metrics (Precision, Recall, F1 Score) for each class, where the weight is the support, or the number of instances, for each class. This average takes into account the class distribution, providing a more realistic view of model performance in imbalanced datasets. Larger classes influence the weighted average more, making it ideal for understanding overall model performance on the dataset as a whole. The weighted average is calculated by summing the metric values $M_i$ for each class, where $M_i$ corresponds to the metric for the $i$-th class, and multiplying each by the support, $n_i$, the number of instances in the $i$-th class. The sum of these weighted values is then divided by the total number of instances $N$ in the dataset.

$$Weighted\ Average = \frac{1}{N} \sum_{i=1}^{c} M_i \times n_i \quad (6)$$

### O. AREA UNDER THE CURVE (AUC) AND RECEIVER OPERATING CHARACTERISTIC CURVE (ROC)

The AUC score evaluates the model's ability to distinguish between COVID and Normal cases. It is derived from the ROC curve, which plots the True Positive Rate (TPR) against the False Positive Rate (FPR) at various classification thresholds. A model achieving an AUC score of 1.0 represents perfect discrimination, while 0.5 represents random guessing. The AUC is calculated as shown in Equation 7. The ROC curve demonstrates a binary classifier's diagnostic ability by plotting the TPR against the FPR as the discrimination threshold varies. TPR and FPR are calculated as shown in Equations 8 and 9.

$$AUC = \int_0^1 TPR(t)\, d(FPR(t)) \quad (7)$$

$$TPR = \frac{TP}{TP+FN} \quad (8)$$

$$FPR = \frac{FP}{FP+TN} \quad (9)$$

### IV. RESULTS

We conducted the experiments using Google Colab, with CPU resources, 51 GB of RAM, and 225.8 GB of disk space. Python 3 and relevant libraries, including Scikit-Learn, Keras, and TensorFlow, were employed to implement the proposed hybrid deep-learning model. We loaded the pre-trained models, namely, VGG16, DenseNet121, and MobileNetV2 architectures from Keras, each initialized with ImageNet weights. We trained the three pretrained learning models and proposed a hybrid deep learning model using the 2108 COVID and non-COVID patient scan images. For model compilation, we employed an Adam optimizer with a learning rate of 1e-4, paired with a binary cross-entropy loss function, which is ideal for binary classification tasks. To enhance the training process, we incorporated several callbacks. Early stopping was utilized to prevent overfitting by monitoring the validation loss and restoring the best weights after a patience period of 5 epochs without improvement. Additionally, we implemented a learning rate reduction strategy that dynamically adjusts the learning rate by a factor of 0.5 when a plateau in validation loss is detected, with a minimum learning rate of 1e-6. Model checkpointing was also integrated to save the best-performing model based on validation loss, ensuring we retain the most effective model after training. The training process was executed on the augmented data, with 20 epochs and a batch size of 8, using the specified callbacks to optimize performance and training efficiency.

We compared the performance of the pre-trained CNN model to that of our proposed model. We evaluated all these models using 373 CT scan images, where 186 images are COVID-infected and 187 are noninfected images, based on various evaluation metrics defined in the methodology section. We generated a confusion report, as shown in Table 4, for each model to evaluate its robustness by determining its accuracy, precision, recall, and f1 score (Table 3). Class level metrics, i.e., COVID and non-COVID confusion reports, are shown in Table 4. The confusion matrix for models is shown in Figure 5. The ROC curve for all models is shown in Figure 6.

TABLE II
PERFORMANCE METRICS OF THE MODELS

| Model | Parameters | Precision (Weighted Avg) | Recall (Weighted Avg) | F1 (Weighted Avg) |
|---|---|---|---|---|
| VGG16 | 88.47% | 88.53% | 88.47% | 88.47% |
| DenseNet121 | 92.76% | 92.79% | 92.76% | 92.76% |

| Model | | Precision | Recall | F1 |
| --- | --- | --- | --- | --- |
| MobileNetV2 | | 94.10% | 94.18% | 94.10% | 94.10% |
| Proposed Hybrid Deep Learning Model | | 98.93% | 98.95% | 98.93% | 98.93% |

TABLE III
CLASS-WISE PERFORMANCE METRICS FOR MODELS

| Model | Class | Precision | Recall | F1 Score |
| --- | --- | --- | --- | --- |
| VGG16 | COVID | 89.94% | 86.56% | 88.22% |
| | non-COVID | 87.11% | 90.37% | 88.71% |
| DenseNet121 | COVID | 93.92% | 91.40% | 92.64% |
| | non-COVID | 91.67% | 94.12% | 92.88% |
| MobileNetV2 | COVID | 96.07% | 91.94% | 93.96% |
| | non-COVID | 92.31% | 96.26% | 94.24% |
| Proposed Hybrid Deep Learning Model | COVID | 100.00% | 97.85% | 98.91% |
| | non-COVID | 97.91% | 100.00% | 98.94% |

The proposed hybrid deep learning model demonstrates superior performance, achieving an accuracy of 98.93%, with weighted average precision, recall, and F1-score all reaching 98.95%. Compared to the individual pre-trained models, MobileNetV2 (94.10%), DenseNet121 (92.76%), and VGG16 (88.47%), the proposed model shows significant improvements. In terms of class-wise performance, the proposed model achieves perfect precision and recall for the non-COVID class (100%) and a remarkable recall of 97.85% for the COVID class, resulting in an overall superior F1 score.

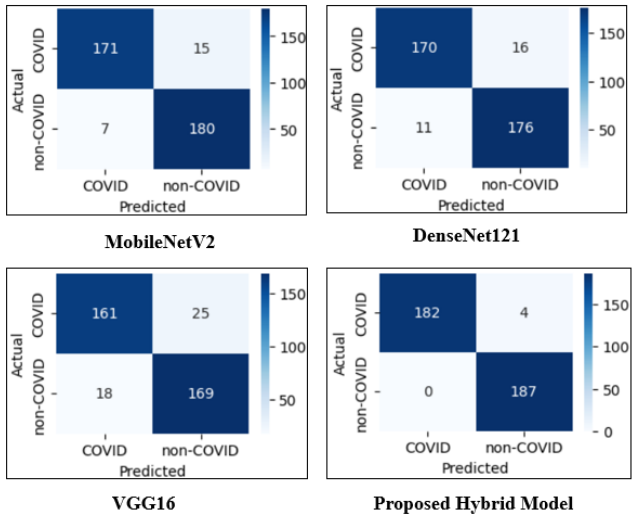

FIGURE 5. Confusion Matrix for Model Evaluation

The confusion matrices presented in Figure 5 compare the performance of the proposed hybrid model against three established models: VGG16, DenseNet121, and MobileNetV2. The results demonstrate a significant improvement in classification accuracy for the proposed model. Specifically, the proposed model achieves almost perfect classification, correctly identifying most of the instances of both COVID-19 and non-COVID cases (182 and 187, respectively), resulting in zero non-COVID misclassifications. In contrast, VGG16 misclassifies 25 COVID-19 cases and 18 non-COVID cases, indicating relatively lower sensitivity and specificity. DenseNet121 performs better, misclassifying 16 COVID-19 cases and 11 non-COVID cases. MobileNetV2 shows further improvement, with only 15 misclassified COVID-19 cases and seven non-COVID cases. These findings underscore the reliability of the proposed model in comparison to existing models.

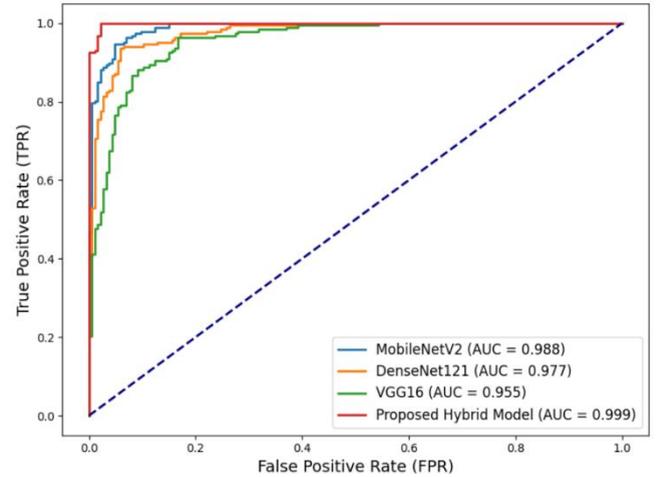

FIGURE 6. ROC Curve for Different Models

The ROC curves depicted in Figure 6 compare the classification performance of the proposed model with three benchmark models: VGG16, DenseNet121, and MobileNetV2. The Area Under the Curve (AUC) values illustrate the superior performance of the proposed model, achieving an AUC of 0.999, indicating near-perfect discrimination between COVID-19 and non-COVID cases. This high AUC value reflects the model's remarkable ability to maintain a high True Positive Rate (TPR) while minimizing the False Positive Rate (FPR). MobileNetV2 demonstrates an AUC of 0.988, followed by DenseNet121 with an AUC of 0.977, and VGG16 with an AUC of 0.955. These results highlight the performance of the proposed model and its potential for real-world deployment in COVID-19 detection scenarios.

## V. Discussions and Limitations

The main reason for the performance differences between VGG16, DenseNet121, and MobileNetV2 can be attributed to the unique strengths and design principles of each architecture. DenseNet121 performed better than VGG16 because it has densely connected layers, which allowed for more efficient information flow and enabled the extraction of intricate patterns from the data. On the other hand, we observed that MobileNetV2 achieved strong performance compared to VGG16 and DenseNet121 due to its efficient use of depth-wise separable convolutions, which reduced computational complexity while maintaining high accuracy, making it ideal for resource-limited environments. Additionally, the use of linear bottleneck layers in MobileNetV2 ensured the preservation of important image features, which is critical for tasks like COVID-19 detection.

The proposed model demonstrates substantial improvements in COVID-19 detection accuracy, precision, recall, and F1 score, as reflected in the results, where the model outperformed individual pre-trained CNNs. These improvements can be attributed to several key factors. First is our image augmentation approach, which, by introducing transformations such as rotations, shifts, and brightness adjustments, simulates real-world variances, helping the model generalize better. This exposure to diverse data variations not only reduces overfitting but also allows the model to learn more discriminative features, contributing to the overall effectiveness of the hybrid model. Another key factor contributing to the improved performance of the proposed model is the combination of features extracted from MobileNetV2, DenseNet121, and VGG16, allowing the model to capture a diverse range of data characteristics. The integration of transfer learning further strengthens the model by leveraging the complementary strengths of each architecture: VGG16 excels in low-level feature extraction, DenseNet121 captures complex high-level patterns, and MobileNetV2 offers efficiency and scalability. This combination reduces overfitting, improves classification accuracy, and proves particularly effective with smaller datasets, as each model contributes its unique strengths to enhance overall performance. Additionally, the use of PCA resulted in improved computational efficiency by refining the feature set, retaining significant variations while reducing redundancy.

The proposed hybrid model demonstrated high performance in COVID-19 detection. However, there are limitations that require further exploration. Our hybrid model has shown that using a transfer learning approach significantly reduces the number of trainable parameters, improving computational efficiency compared to existing ensemble approaches for COVID-19 detection. However, it also required more computational resources compared to using an individual transfer learning model, such as DenseNet121 or MobileNetV2. In future work, we plan to focus on optimizing efficiency without reducing the model's performance, making it more suitable for deployment in resource-constrained environments. While our study successfully achieves its primary goal of enhancing COVID detection, future research should evaluate the performance of the proposed ensemble approach on larger and more diverse datasets to assess its scalability and adaptability. Additionally, exploring alternative data augmentation methods, advanced optimization strategies, and hyperparameter tuning could further improve the model's robustness and accuracy. Since the proposed approach, like many deep learning models, operates as a black box with limited interpretability, incorporating explainable AI techniques in future work could enhance transparency and trust in its decision-making process, making it more suitable for clinical use.

## VI. CONCLUSION

The results obtained from our study demonstrate that the proposed hybrid deep learning model significantly improves COVID-19 detection from CT scan images, achieving an accuracy of 98.93%. By combining transfer learning from three pre-trained CNNs, namely VGG16, DenseNet121, and MobileNetV2, with Principal Component Analysis (PCA) and Support Vector Classification (SVC), the proposed model outperforms individual CNN models in terms of accuracy, precision, recall, F1 scores, and ROC curve performance. The comparative analysis confirms that the proposed model exhibits superior performance with minimal misclassifications. Our proposed approach can be extended to detect other diseases using CT scan images, leveraging the power of deep learning and transfer learning techniques. We plan to focus our future research on exploring alternative feature extraction methods, fusion techniques, and advanced image processing or augmentation strategies, which could further enhance the performance of the proposed model. Furthermore, we plan to optimize the computational efficiency of the proposed model in the future to enable its deployment in real-world clinical settings, particularly in resource-constrained environments like mobile or edge devices, with faster inference.

**SURESH BABU NETTUR** has received his Master of Science (M.S) degree from the Birla Institute of Technology and Science (BITS), Pilani, Rajasthan, India, and his Bachelor of Technology degree in Computer Science and Engineering from Nagarjuna University, Guntur, India. With over two decades of expertise, he has established himself as a thought leader in artificial intelligence, deep learning, and machine learning solutions. His work spans the development of scalable and intelligent systems across industries such as healthcare, finance, telecom, and manufacturing. Suresh has been at the forefront of integrating AI-driven innovations into real-world applications, leveraging cutting-edge technologies such as OpenAI models, GitHub Copilot, and custom deep learning architectures to deliver transformative solutions. His contributions include designing and implementing advanced machine learning models, optimizing deep learning architectures for resource-constrained environments, and integrating AI solutions into software development and testing pipelines. With significant experience in cross-functional team leadership and managing onsite-offshore collaboration models, he has successfully delivered AI-powered applications across cloud platforms like AWS.

Suresh is passionate about applying AI to solve complex problems in healthcare and finance, including predictive analytics, automation, and intelligent decision-making systems. He is proficient in Agile methodologies, Test-Driven Development (TDD), and service-oriented architectures (SOA), ensuring seamless integration of AI and machine learning into software systems. As an advocate for innovative AI applications, Suresh is committed to advancing the field through sustainable and impactful solutions that redefine industry standards and improve quality of life.

**SHANTHI KARPURAPU** received the Bachelor of Technology degree in chemical engineering from Osmania University, Hyderabad, India and the Masters technology degree in chemical engineering from Institute of Chemical Technology, Mumbai, India.
She has over a decade of experience leading, designing, and developing test automation solutions for various platforms across healthcare, banking, and manufacturing industries using Agile and Waterfall methodologies. She is experienced in building reusable and extendable automation frameworks for web applications, REST, SOAP, and microservices. She is a strong follower of the shift-left testing approach, a certified AWS Cloud practitioner, and a machine learning specialist. She is passionate about utilizing AI-related technologies in software testing and the healthcare industry.

**UNNATI NETTUR** currently pursuing an undergraduate degree in Computer Science at Virginia Tech, Blacksburg, VA, USA. She possesses an avid curiosity about the constantly evolving field of technology and software development, with a particular interest in Artificial Intelligence. She is passionate about gaining experience in building innovative and creative solutions for current issues in the field of software engineering.

**LIKHIT SAGAR GAJJA** pursuing a Computer Science Bachelor's degree at BML Munjal University, Haryana, INDIA. He is evident in showing his passion for the dynamic field of technology and software development. His specific interests include Artificial Intelligence, Prompt Engineering, and Game Designing technologies, highlighting his dedication to obtaining hands-on experience and developing innovative solutions for real-time issues in software engineering.

**SRAVANTHY MYNENI** earned master of science in information technology and management from Illinois institute of technology, Chicago, Illinois in 2017 and Bachelor's degree in computer science in 2013. She is currently working as an engineer focused on data engineering and analysis. She has 8+ years of experience in designing, building and deploying data centric solutions using Agile and Waterfall methodologies. She is enthusiastic about data analysis, data engineering and AI application to provide solutions for real world problems.

**AKHIL DUSI** currently pursuing Masters of Information Sciences at University of Indiana Tech, Indiana, USA. He is a passionate researcher and developer with a diverse background in software development, cybersecurity, and emerging technologies. He has a proven ability to deliver innovative and practical solutions. His work includes developing web and mobile-based applications, conducting vulnerability assessments and penetration testing, and leveraging cloud platforms for efficient infrastructure management. He is certified in cybersecurity and machine learning, reflecting a strong commitment to continuous learning and staying at the forefront of technological advancements. His research interests focus on artificial intelligence, IoT, and secure system design, with a vision to drive impactful innovations.

**LALITHYA POSHAM** is an MBBS graduate from Nanjing Medical University. She has a strong passion for advancing clinical research and improving patient outcomes. With a solid foundation in medical education, she is particularly interested in exploring innovative diagnostic approaches and aims to integrate clinical expertise with research to drive improvements in healthcare systems.